\documentclass[prl,reprint,showpacs,floatfix,nofootinbib]{revtex4-1}
\usepackage{epsfig,graphics,float,amsmath,mathtools}
\usepackage[colorlinks=true,citecolor=blue]{hyperref}
\usepackage{microtype,setspace,siunitx,physics,epsfig,graphicx}
\usepackage[caption=false]{subfig}
\usepackage{xcolor}
\usepackage[export]{adjustbox}
\usepackage{nameref,physics}
\usepackage{blindtext}
\usepackage[english]{babel}
\usepackage[normalem]{ulem}
\useunder{\uline}{\ul}{}
\usepackage{graphics}

\begin{document}
\title{End-to-End Machine Learning for Experimental Physics: Using Simulated Data to Train a Neural Network for Object Detection in Video Microscopy}
\date{\today}
\author{Eric~N.~Minor}
\author{Stian~D.~Howard}
\author{Adam~A.~S.~Green}
\author{Cheol~S.~Park}
\author{Noel~A.~Clark}
\affiliation{Department of Physics and Soft Materials Research Center, University of Colorado, Boulder, Colorado, 80309, USA}

\begin{abstract}
    
    We demonstrate a method for training a convolutional neural network with simulated images for usage on real-world experimental data. Modern machine learning methods require large, robust training data sets to generate accurate predictions. Generating these large training sets requires a significant up-front time investment that is often impractical for small-scale applications. Here we demonstrate a `full-stack' computational solution, where the training data set is generated on-the-fly using a noise injection process to produce simulated data characteristic of the experimental system.
    
    
   We demonstrate the power of this full-stack
   approach by applying it to the study of topological defect annihilation in systems of liquid crystal freely-suspended films. This specific experimental system requires accurate observations of both the spatial distribution of the defects and the total number of defects, making it an ideal system for testing the robustness of the trained network. The fully trained network was found to be comparable in accuracy to human hand-annotation, with four-orders of magnitude improvement in time efficiency.  
    
\end{abstract}

\maketitle

\section{Introduction}
Current generation physics experiments often produce data of such complexity and volume that belie efficient analysis by classical algorithms\cite{Adam-BourdariosHiggsMachineLearning2015,radovic_machine_2018}. The modern renaissance in machine learning\cite{dey_machine_2016,BishopPatternRecognitionMachine2006,Tanartificialintelligencerenaissance2018} provides a potentially superior method to analyze such complex data. Indeed, machine learning methods have been successfully implemented in many scientific disciplines; from  high-energy physics\cite{baldi_searching_2014,AlbertssonMachineLearningHigh2018}, and condensed matter physics\cite{deng_machine_2017,carrasquilla_machine_2017,beach_machine_2018,wang_machine_2017,walters_machine_2019} to biological systems\cite{tarca_machine_2007}. These machine-learning algorithms often preform more robustly than previous state-of-the-art solutions\cite{al-jarrah_efficient_2015}.


A characteristic realization of a data set with both volume and complexity is that of an image sequence produced by video microscopy, a ubiquitous method of dynamical systems analysis that occurs in fields as disparate as biological systems\cite{kner_super-resolution_2009,Langebasicsimmunofluorescencevideomicroscopy1995} and hydrodynamics\cite{crocker_methods_1996,kellay2017}.

The analysis of video microscopy has long been stymied by the inherent difficulty of extracting quantitative information from images\cite{baumgartl_limits_2005}.  Except for very simple tasks, the classical algorithms to extract quantitative information from these videos require very specific conditions\cite{conte_thirty_2004}, requiring significant pre-processing and strict human oversight. Often these ideal conditions cannot be realized in an experimental setting, necessitating the bespoke analysis of individual image frames, where manual extraction of quantitative data is required. This results in a significant bottleneck in experimental analysis.

One common objective in the analysis of video microscopy is the extraction of the spatial position of a defined target. Machine learning methods have already been deployed to great effect in this regard, where they prove capable of both spatially labelling and categorizing human-defined objects\cite{ErhanScalableObjectDetection2014,BlaschkoLearningLocalizeObjects2008}. However, these models are reliant on large, previously analyzed training sets, where the target has already been identified and annotated. The generation of these training sets requires a large, upfront time investment that make them impractical for small-scale applications.

Here we report on an `end-to-end' method, where the training data and annotations, the list of spatial coordinates of targets, are procedurally generated through computer simulation, allowing for fast deployment of machine learning methods to small-scale applications.


We demonstrate the robustness of our approach through the analysis of defect-defect interactions in freely-suspended films of smectic-C liquid crystal.

\subsection{Background}
A smectic phase is a liquid crystalline mesophase composed of elongated molecules, with general orientational order between between the molecules and crystalline order along one axis.
The crystalline order segregates the phase into stacked sheets of molecules that can flow freely in the plane, making smectic liquid crystals ideal realizations of two-dimensional hydrodynamic systems. Additionally, the molecules in the phase can be oriented co-linearly with the smectic-plane normal vector (SmA) or can be tilted with respect to the smectic-plane normal vector (SmC). In the latter case, the molecular tilt breaks the isotropic nature, giving the SmC phase a rich topological structure. 

To first order, the Frank free energy that describes a single smectic-c layer is well approximated by the continuous XY model, which supports as ground-state solutions stable topological defects. The theoretical\cite{yurke_coarsening_1993, svensek_hydrodynamics_2002, svensek_hydrodynamics_2003,radzihovsky_anomalous_2015, pleiner_dynamics_1988}and experimental\cite{pargellis_defect_1992, pargellis_planar_1994, oswald_nematic_2005,stannarius_defect_2016} dynamics of defects in liquid crystal systems has been studied since the early 90's. However, it is an open question how well the non-hydrodynamic XY model describes the interaction of these topological defects in fluidic systems of liquid crystal materials.

Direct tests of the XY model could be made by observing the total number of and nearest-neighbor distance of defects in the coarsening dynamics of a quenched SmC film, but, as there currently exists no robust way to spatially track or label the defects in these textures, this analysis must be done manually-- severely limiting the temporal resolution.

Machine-learning methods have already been successfully deployed in studies of the XY model, with previous work demonstrating the viability of using basic neural networks to identify whether a given simulated data set contains a topological defect\cite{walters_machine_2019}. However, the work was focused on simulated system states, consisting of molecule locations and orientations, rather than experimental image analysis. Furthermore, the algorithm that was utilized is purely for classification and did not give defect counts or locations, limiting uses in experimental data analysis.

\section{Experimental System}


In order to confirm the veracity of our system, we collected physical data from a typical topological defect experiment.
To generate data for training the machine learning system, we used a simulation to generate perfectly annotated images and then ran those images through a data enhancement pipeline.




\subsection{Experimental Defect Data}

\begin{figure}
  \includegraphics[width=\linewidth]{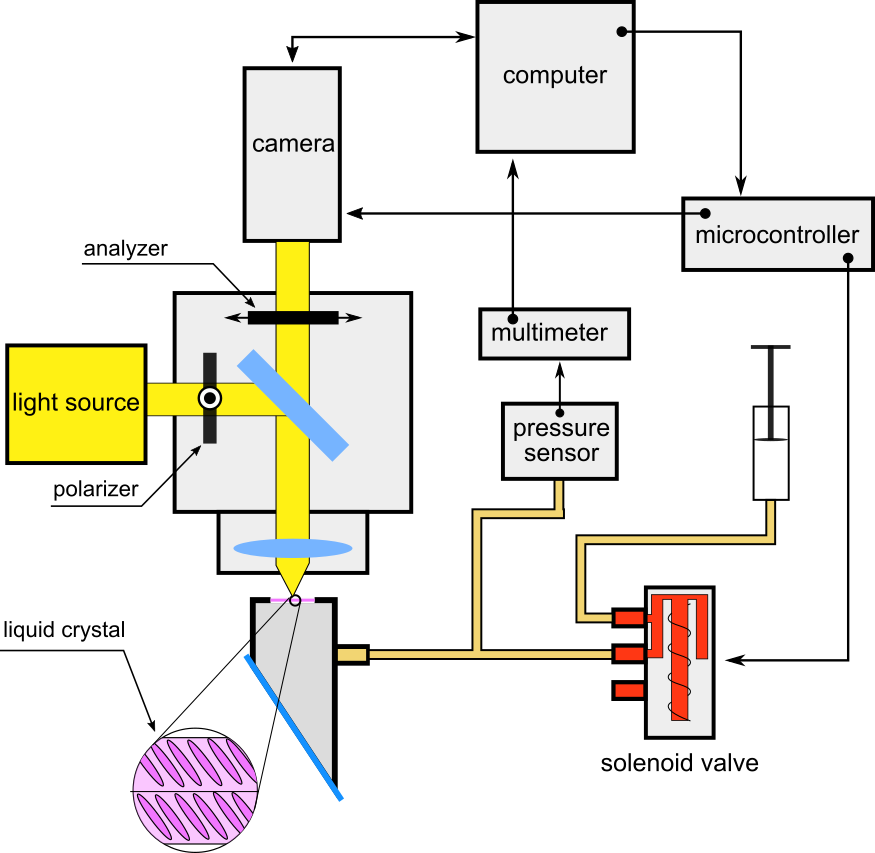}
  \caption{Schematic of mechanical quench experiment. The syringe is coupled to a translation stage, allowing for fine control.
When depressed, the compression increases the pressure in the quench chamber,
causing the film to bulge outwards. The pressure is monitored by a pressure
sensor (Honeywell SDX010IND4) which outputs a proportional voltage read by a
digital multimeter connected serially to a computer. The user can enter a
target threshold pressure in a program running on the computer, which
automatically triggers a quench when said threshold is reached. The quench is
triggered by sending a digital signal to a connected microcontrolled (Arduino
Uno) which first instructs the high-speed (Phantom V12.2) camera to begin
recording and then opens the solenoid valve, venting the system to atmosphere.
The resulting dynamics on the film are captured by the camera as a video, which
is then transferred to the computer for further analysis.}
  \label{fig:Setup Sketch}
\end{figure}

At the center of our experimental setup, shown in Figure~\ref{fig:Setup Sketch}, is a pressure chamber with an open aperture on the top to draw a film over. Air is pumped into the chamber, increasing pressure and causing the film to bulge outward. A valve on the tube is then opened to the atmosphere, rapidly equalizing the pressure in the chamber. The valve is controlled by a computer program that, upon reaching a predetermined pressure differential between the chamber and atmosphere, opens the valve, triggers the high-speed camera recording, and starts saving pressure readings.  The collapse of the film results in a mechanical quench, which creates a high-energy state resulting in a large density of defects in the film which rapidly annihilate.

\begin{figure}[h!]
  \includegraphics[width=\linewidth]{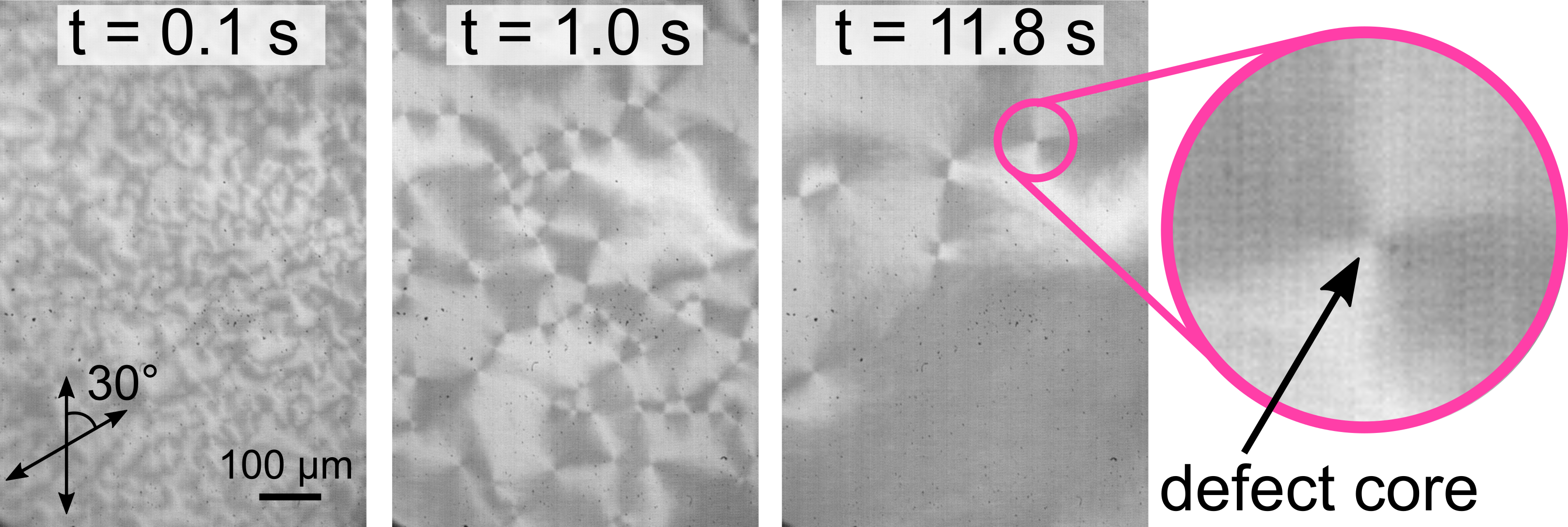}
  \caption{Experimental data showing time evolution of a film with topological defects viewed under reflection microscopy. Images are labelled with the time elapsed since quenching. The topological defect appears as a `bowtie' structure.}
  \label{fig:frames}
\end{figure}

SmC liquid crystal defects can be visualized with partially or fully crossed polarizers. In our setup, polarized light is shined perpendicularly onto the film. The reflected light is collected into a microscope where it passes through a partially crossed polarizer. Because of the birefringent nature of SmC liquid crystals, when viewed under crossed polarizers the orientation of the molecule is mapped to a reflected intensity as $I \propto \sin(2\theta)^2$, where $\theta$ is the angle between the in-plane projection of the molecule (refered to as the c-director) and the polarizer. However, working with fully crossed polarizers dramatically decreases the reflected light, acting as a significant limiting factor for the exposure time needed to get viable images. Therefore, we work in a regime of decrossed polarization, which reflects more light. In this regime, the reflected intensity is well approximated by $I \propto \cos(2\theta)$, giving a characteristic `bowtie' structure, as seen in Figure~\ref{fig:frames}\cite{chattham_triclinic_2010}. A high speed camera (Phantom V12.1) records the reflected light in gray-scale at 500 frames per second with an exposure time of 1900 $\mu$s, allowing us to directly view the coarsening dynamics of the film.

Each video lasts 12.2 seconds, capturing the entirety of the short term dynamics. The images, with 1104x800 resolution and 12 bit pixel precision, allow for high contrast to be gained in post processing.
We used PM2 \cite{harth_episodes_2016} to form a film that exists in a Smectic C phase at room temperature. Figure~\ref{fig:frames} provides snapshots of the data collected over a range of times.

\subsection{Simulation Data}

Two methods were used to generate images for training the machine learning models to detect topological defects. 
The first method procedurally generates textures by linearly adding a random number of defects at random locations to an initially aligned XY grid. In the XY model, a stable, zero-temperature solution of the XY Hamiltonian is given by the plus/minus defect director configuration:
\begin{equation}
\varphi(x,y) = \pm \arctan{\left(\frac{y-y_0}{x-x_0}\right)} + \phi_0, 
\end{equation}
where $\phi_0$ is a phase offset which was also randomized, and $(x_0,y_0)$ is the location of the defect.

\begin{figure}
  \includegraphics[width=\linewidth]{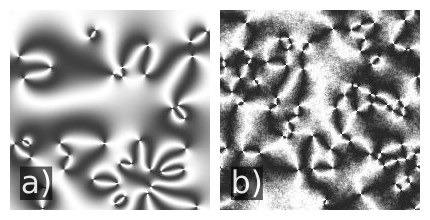}
  \caption{Typical image from random-defect simulation (a) and thermal-defect simulation (b). }
  \label{fig:RandomDefect}
\end{figure}

In this way, arbitrarily complex defect configurations can be generated. The c-director can be mapped to a scalar intensity value using the Schlieren mapping $I = \cos(2\varphi)$, giving a reasonable facsimile of experimental observations of defect configurations in freely-suspended films. Because the method involves linear superpositions of zero-temperature solutions, it produces very clean images that are free of thermal noise, as seen in Figure \ref{fig:RandomDefect} (a). 
 This simulation method will be referred to as the `random defect' model.

The second method is based on directly simulating the dynamics of the XY model at a finite temperature as it evolves from a high-density defect configuration\cite{loft_numerical_1987, yurke_coarsening_1993, jelic_quench_2011}. The angle of the c-director at each lattice site $i$ evolves according to the discretized Ginzburg-Landau model through the Euler update scheme as:
\begin{equation}
    \varphi_i(t+\Delta t) = \varphi_i(t) -\Delta t \left( \eta_i(t) + \kappa \sum_{\langle i,j \rangle} \sin(\varphi_i(t) - \varphi_j(t)) \right)
\end{equation}
where $\kappa$ is a visco-elastic constant, and $\eta_i$ is a random number with moments that correspond to the temperature through the fluctuation-dissipation theorem:
\begin{equation}
    \langle \eta_i(t) \eta_j(t') \rangle  = 2 T \delta_{i,j} \delta(t-t')
\end{equation}
Using the Chester-Tobochnik method\cite{tobochnik_monte_1979}, the defect locations can be extracted by calculating the winding number around each lattice plaquette. This is done by computing the successive differences in angles as the plaquette is circumnavigated, defined as: $\Delta \varphi = 2 \pi n_i+ \theta$, where $\theta$ is restricted to the range $(-\pi,\pi)$. The vorticity of the square is equal to the sum of the $n_i$'s, and can be either positive or negative. In this way, the locations and number of the defects for each time-step in the system can be extracted in an automated annotation process. This method of directly simulating the XY model is capable of producing a wide variety of textures with different amounts of director fluctuations, shown in Figure~\ref{fig:RandomDefect} (b). This method will be refered to as the `thermal-defect' model. The inclusion of temperature means these simulations more strongly conform to the experimental observations, making the thermal-defect model the preferred method.

\section{Topological Defect Tracking}

In order to make a system viable for usage with real experiments, we developed a pipeline that makes use of modern deep-learning object detection and image enhancement techniques to train a model capable determining both the location and count of objects in an image. 

\subsection{Pipeline and Image Enhancement Motivation}
For object detection we used darkflow\cite{trinh_darkflow_2018}, a TensorFlow implementation of the YOLOv2 algorithm \cite{redmon_yolo9000:_2016} that offers improved performance when compared to the original YOLO algorithm\cite{redmon_you_2015}. YOLOv2 learns to perform both region proposal and region classification using the darknet-19 architecture. Training the region proposal mechanism is important for defect identification as detection algorithms that rely on traditional heuristic searches, such as R-CNN\cite{girshick_rich_2013} and its successors, Fast R-CNN\cite{girshick_fast_2015} and Faster R-CNN\cite{ren_faster_2017}, would likely fail to identify defects as objects. Training and testing darkflow on a set of 100 simulated 200x200 images, with each image containing 20 defects, showed that this network is viable for detecting the locations of defects in simulation data. However, data enhancement techniques are necessary to make a network trained on simulated images viable for application to real data.

Machine learning algorithms, by nature, optimize themselves to perform as well as their architecture allows on the given training data. While this can lead to highly effective systems, it is the primary reason why training on a set of simulated data often makes the final model non-viable in the real world; simulated data is highly predictable and clean while real-world data can have significant noise and variance in how key objects appear. By training on the simulated data, the system will over-fit\cite{lawrence_lessons_1997} \cite{lever_model_2016} on the very specific shapes, textures, and gradients produced by the simulation. 
Our solution for training a model on simulated data to analyze real data is to introduce various artifacts that mimic real-world inaccuracies into the simulated images.

\subsection{Standardization and Simulated Image Enhancement}

The first issue that needs to be dealt with is lighting and contrast. In simulated defect images, the intensity of a pixel ranges from perfectly black to perfectly white depending on the director orientation, maximizing the gradients and contrast in the image. The mean intensity of the image will also generally be around 0.5 on a scale from 0 (black) to 1 (white) since there is no offset to the image brightness. When using an experimental image, the difference in brightness between perfectly aligned and perfectly misaligned directors is much smaller than the full dynamic range of the image, which causes smaller gradients. The average brightness of the experimental data is rarely 0.5, so what constitutes bright and dark pixels is more complex than just the intensity of the pixel. To make the simulation and experimental images as similar as possible in regards to average intensity and dynamic range, a variant of the basic feature standardization procedure\cite{aksoy_feature_2001} is used. Each pixel's intensity value is set according to the feature standardization formula

\begin{equation}
 x' = \frac{x - \Bar{x}}{6\sigma} +0.5
\end{equation}
where x' is the output pixel intensity, x is the input intensity of each pixel and $\sigma$ is the standard deviation of the global image pixel intensities. The output images have a mean pixel intensity of 0.5 and a dynamic range of six standard deviations. This procedure reasonably standardizes the lighting and contrast of the images regardless of the actual lighting and camera conditions, providing consistency across multiple data sets.

Adding imperfections to the simulated images emulates the experimental data and improves the robustness\cite{goodfellow_explaining_2014, bishop_training_1995} of the neural network model. Gaussian blurring, Fourier noise, randomized image variance, randomized lighting boundaries, and arbitrary objects each target identified inconsistencies between simulation and experimental data. The alterations increase the image variety in our training data-set and teach the model that these imperfections are to be ignored when attempting to detect defects.

\begin{figure}
  \includegraphics[width=\linewidth]{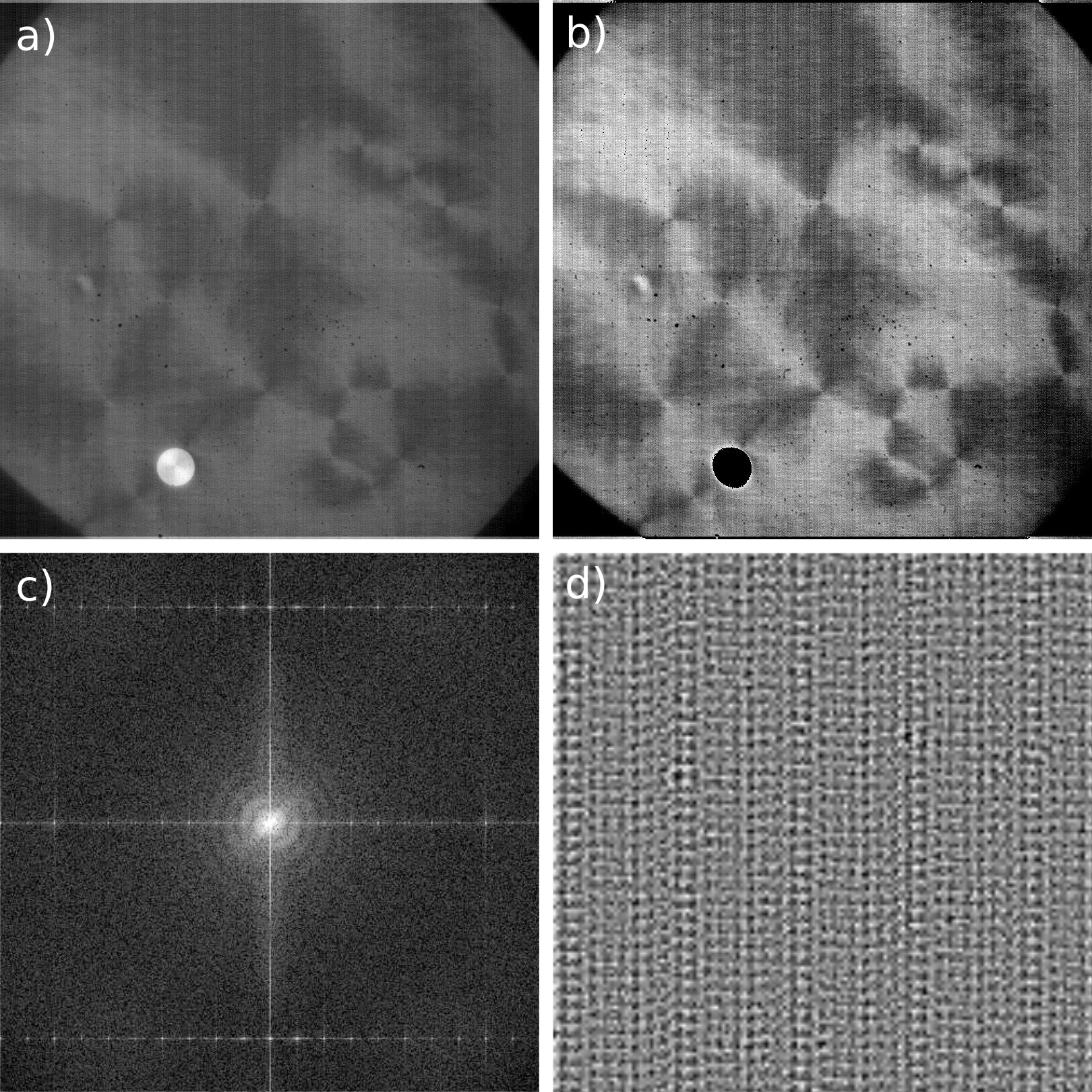}
  \caption{Image standardization and periodic noise extraction. a) Raw image with an island. b) Standardized image. c) Magnified Fourier transform of standardized image. d) Magnified noise extracted from the image.}
  \label{fig:Standardization and Noise}
\end{figure}

\begin{figure}[h!]
  \includegraphics[width=\linewidth*3/4]{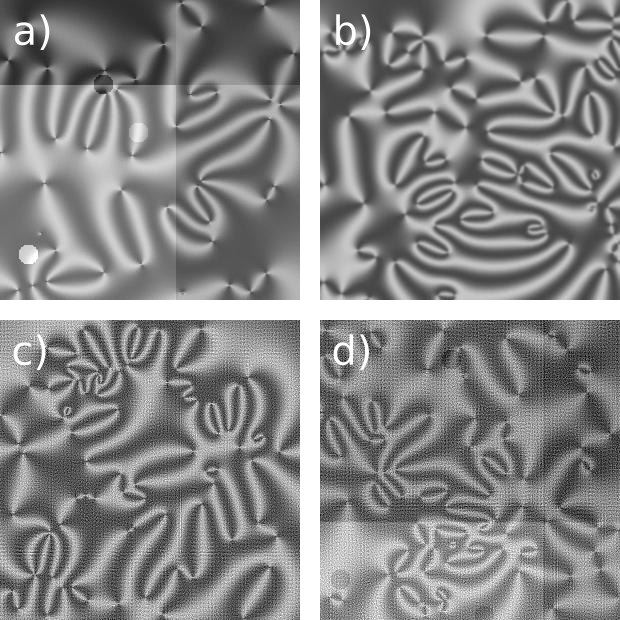}
  \caption{Simulated images enhanced with various types of noise. (a) Image with dramatic lighting shifts and added circles (islands) (b) Image with Gaussian blurring (c) Image with Fourier noise (d) Image with all three}
  \label{fig:Image Enhancement}
\end{figure}

Due to the relatively low lighting of the experimental images, the camera read noise, generated by the camera hardware when reading information from the CCD (Charge-Coupled Device), is significant relative to the signal size. Applying a 2-D discrete Fourier transform, the composition of the image is extracted in the frequency domain\cite{kaur_periodic_2014}, shown in Figure~\ref{fig:Standardization and Noise}, where the periodic read noise appears as regularly-spaced lines. This characteristic camera noise is added as low-frequency noise to the simulated data set to increase similarity to experimental data, shown in Figure~\ref{fig:Image Enhancement} c.

When collecting experimental data, it is rare that perfect focus is consistently achieved. In the quenching experiment, there are additional film fluctuations as the pressure on the two sides of the film equalize, resulting in a shifting focus point as the film moves. The lighting conditions can change significantly for different experimental setups. In particular, film thickness and camera settings have an impact on the intensities of light captured by the camera. Randomized Gaussian blurring emulates the non-perfect and variable focus of the experimental data, shown in Figure~\ref{fig:Image Enhancement} (b). The overall brightness and dynamic range of the simulated images is randomized to prevent the model from being dependant on specific light intensities or gradient magnitudes unique to the simulation.

The final additions to the simulated images are randomized lighting domains and circular artifacts. Observing the pattern of defect detections from previous models, it was discovered that detections would be made along lines where the lighting abruptly shifts. The microscope aperture and the boundaries between CCD sectors, which consistently read slightly different pixel intensities, generate the light shifts. To prevent this, the simulation images were broken into four quadrants of randomized size, with each quadrant having slightly different brightness. False detections were also made around the boundaries of islands, which are regions in liquid crystal films with additional layers of material. To prevent this, circles of random brightness were added to the simulation data to provide neutral examples \cite{koppel_importance_2006} of non-defect objects that should not affect detections, shown in Figure~\ref{fig:Image Enhancement}(c).

\subsection{Effects of Simulated Image Enhancements}

To evaluate the effectiveness of each component in the pipeline, several models were trained on simulated images enhanced by various combinations of pipeline components. The models were validated using a hand-annotated set of experimental images to determine how well they performed on real data relative to human performance. The efficacy of the machine learning can be quantified through the \textit{precision} and the \textit{recall}. Precision describes the accuracy of object detection. Recall describes how many of the objects in the image we detect. Rigoursly, these quantities are defined as follows, where $TP$ represents the true positive detections, $FP$ represents the false positive detections, and $FN$ represents the false negative detections.
\begin{equation}
Precision = \frac{TP}{TP+FP}
\end{equation}
\begin{equation}
Recall = \frac{TP}{TP+FN}
\end{equation} 

The model provides a confidence score for each detection. A threshold is set to drop low confidence detections. A low threshold will increase recall while lowering precision; a high threshold will have the opposite effect by only using the few detections that have a high likelihood of being correct. mAP and peak F1 scores are used to characterize the model across all thresholds. AP is the Average Precision across all recall scores and is a general measure for the effectiveness of the model across all thresholds. mAP is the mean AP across all detected classes, however since we are only training to identify defects, mAP is effectively equivalent to AP. The F1 score is the harmonic mean of the precision and recall, and it provides an overall `goodness' measure in regards to recall and precision at a specific threshold. We recorded the maximum F1 score of each model to provide a measure of the peak model performance when choosing an ideal threshold. The mAP\cite{everingham_pascal_2010} score can be thought of as measuring average performance without setting a minimum confidence threshold while peak F1\cite{chinchor_muc-4_1992} score measures the highest obtained performance over all thresholds. Using $p(r)$ to represent the precision at a given recall value and $n$ to represent the total number of detections, AP and F1 can be defined as follows:

\begin{figure}
  \includegraphics[width=\linewidth]{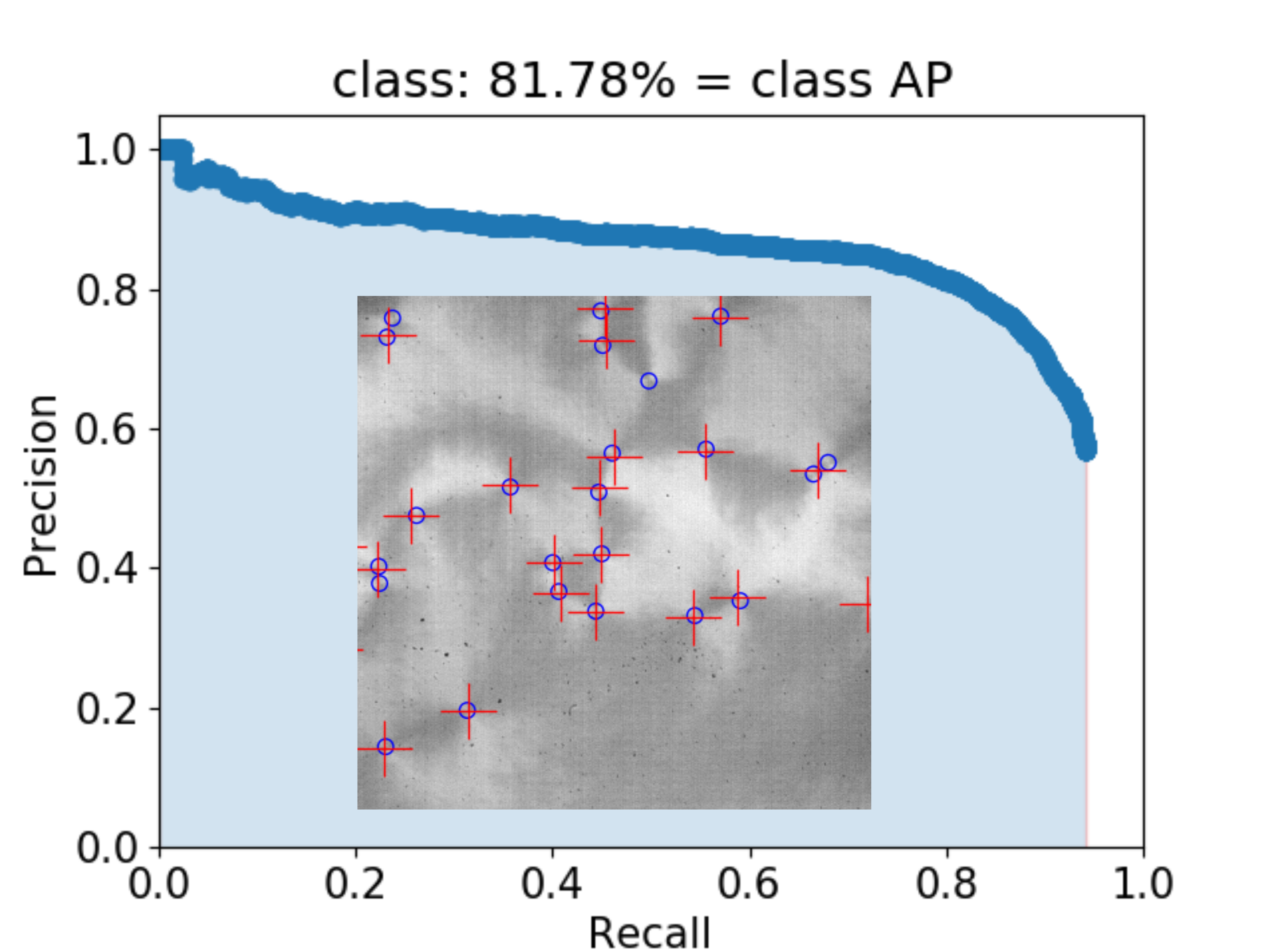}
  \caption{Precision-Recall plot for the highest scoring model, with inset showing a characteristic annotated video still, with blue circles showing hand labelled defects and red crosses showing YOLO detections. mAP score is the percentage of area shaded between (0,0) and (1,1). Conceptually, this means that an average precision of 81.78\% was achieved across all possible recall values}
  \label{fig:mAP score}
\end{figure}

\begin{equation}
F1 = 2*\frac{Precision * Recall}{Precision + Recall}
\end{equation}

$$ AP = \int_0^1 \!p(r) \, \mathrm{d}r $$
however, in order to use AP for real data the formula must be discretized into
\begin{equation}
AP = \sum_{r=1}^{n} p_{interp}(r)
\end{equation}
where
$$ p_{interp}(r) = max_{\bar{r}:\bar{r}>r} p(\bar{r})  $$
Using the max function smooths the AP curve, preventing the local dips at each false detection from affecting the global score. In Fig. \ref{fig:mAP score}, the mAP score is represented by the area of the shaded region.

Simulated training images enhanced with only blurring, random islands, random lighting quadrants, or randomized image brightness and contrast produced models that performed poorly on experimental images. On validation, these models received mAP scores of $<30\%$. Simulated images enhanced only with noise extracted using the Fourier transform produced a viable model, however the mAP score only reached $53.3\%$.

Adding multiple types of noise to the simulated images produced greatly improved models. Combining all types of artifacts produced a model with a mAP score of $59.0\%$, however this was outperformed by a model trained using only Fourier noise and randomized blurring which achieved a mAP score of $69.3\%$. Adding all forms of noise at a lower intensity to the simulated images produced a model with a mAP score of $74.9\%$ [Fig. \ref{fig:Image Enhancement} d]. This suggests that a balance must be struck between making modifications and maintaining enough clarity to identify objects when training.

The original YOLOv2 \cite{redmon_yolo9000:_2016} paper reported an average mAP score of $73.4\%$ when tested using the Pascal VOC2012 test set, which puts the detection accuracy of our model trained with simulated images on par with models trained using real images. This supports the viability of training YOLO object detection models with simulated data for use on experimental data.

\subsection{Improvements Using the XY Model}
Models trained with data produced from the simulation using a Landau-Ginzberg implementation of the XY model yielded improved accuracy. Landau-Ginzberg simulations provide thermal noise, as we see in our real data, and emulate natural defect systems. With no image modifications, a model trained on simulated images from the XY simulation attained a $47.4\%$ mAP score, a significant improvement over the 2\% achieved with the model trained on the raw random defects data. 

\begin{table}[h!]
\resizebox{\columnwidth}{!}{%
\begin{tabular}{lll}
\hline
\multicolumn{3}{l}{Table 1}                                                                                                                   \\
\multicolumn{3}{l}{\textit{\begin{tabular}[c]{@{}l@{}}Scoring of Models Trained\\   on Simulated Images with Various Artifacts\end{tabular}}} \\ \hline
{\ul Artifacts Added}                               & {\ul Peak F1}                      & {\ul mAP}                                    \\
LG, L-FN, L-RV, L-RB, L-GB, L-DC, RD, LR            & 0.811                                    & 0.818                                        \\
LG, L-FN, L-RV, L-RB, L-GB, L-DC, RD                & 0.817                                    & 0.808                                        \\
LG, FN, RV, RB, GB, DC, RD                          & 0.806                                    & 0.783                                        \\
L-FN, L-RV, L-RB, L-GB, L-DC, RD                    & 0.754                                    & 0.749                                        \\
FN, RB                                              & 0.744                                    & 0.738                                        \\
FN, RV                                              & 0.726                                    & 0.725                                        \\
FN, RV, RB, GB, DC, LT                              & 0.740                                    & 0.700                                        \\
FN, GB                                              & 0.707                                    & 0.693                                        \\
FN, RV, RB, GB, DC, RD                              & 0.683                                    & 0.663                                        \\
FN, RV, H-RB, GB, DC, RD                            & 0.661                                    & 0.643                                        \\
FN, RV, RB, H-GB, DC, RD                            & 0.635                                    & 0.626                                        \\
FN, RV, RB, GB, H-DC, RD                            & 0.632                                    & 0.620                                        \\
H-FN, RV, RB, GB, DC, RD                            & 0.640                                    & 0.617                                        \\
FN, RV, RB, GB, DC                                  & 0.613                                    & 0.590                                        \\
FN, H-RV, RB, GB, DC, RD                            & 0.587                                    & 0.579                                        \\
FN                                                  & 0.569                                    & 0.533
                                     \\
LG                                                  & 0.513                                    & 0.474 
\\
RB                                                  & 0.442                                    & 0.270                                        \\
RV                                                  & 0.423                                    & 0.260                                        \\
GB                                                  & 0.296                                    & 0.116
\\
Raw Random Defect Sim & 0.099 & 0.020

\\ \hline
                          
\end{tabular}%
}

\resizebox{\columnwidth}{!}{%
\begin{tabular}{lll}
\multicolumn{3}{l}{\textit{Key}}                                                            \\
FN: Fourier Noise       & GB: Random Gaussian Blurring  & LT: Longer Training Time          \\
RV: Random Variance     & DC: Randomized Decross Angle  & L-XX: Lowered randomization of XX \\
RB: Random Boundaries   & RD: Randomized Defect Number  & H-XX: Higher randomization of XX  \\
LR: Long Simulation Run & LG: Used LandauGin Simulation &                                  
\end{tabular}%
}
\end{table}

The Landau-Ginzberg simulations are highly time-dependent with defect counts following a power law. This means that a linear reduction in defect number requires an exponential amount of time. If we train a model with only early time simulation images, where there is a high density of defects in the training data, the model will perform worse on images with a lower defect density. As such, the best performing models require long simulation runs to generate training data with a wide variety of defect densities. 

Similar to the random defect simulation data, the best results are attained with a lower intensity of many different forms of noise, achieving a mAP score of $81.8\%$. A full account of the artifacts added when training models and the evaluation metrics for each model can be found in Table 1.

\subsection{Model Applications}

When applying the model to data, a threshold needs to be set to eliminate low-confidence detections. To maximize the trade-off between precision and recall, the threshold corresponding to the model's peak F1 score is used. To evaluate the applied performance of the system, we use the top-scoring model that employed Landau-Ginzberg simulation and moderate levels of image enhancement.

\begin{figure}[h!]
  \includegraphics[width=.8\textwidth]{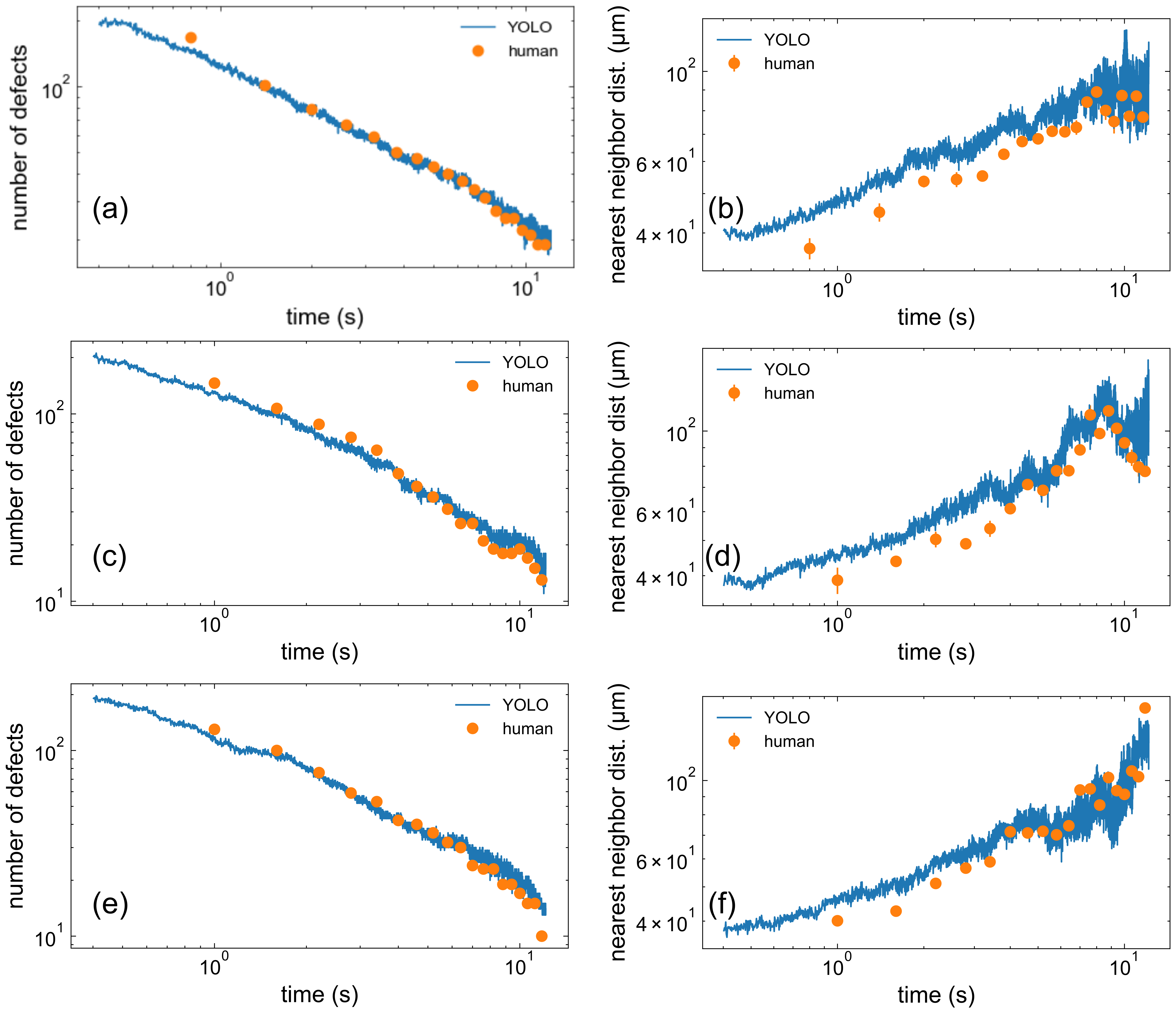}
  \caption{Validating YOLO defect detections. (a,c,e) show number of defects vs. time for three separate videos, comparing YOLO results with hand annotated results. (b,d,f) show defect nearest neighbour distance vs. time for three separate videos, comparing YOLO results with hand annotated results.}
  \label{fig:HumanVMachine}
\end{figure}

A straightforward application of the model is counting the defects per frame in a video. Accurately resolving the defect number as a function of time would allow direct experimental probes of the applicability of the XY model in these systems of SmC liquid crystals. The model results, as seen in Figure~\ref{fig:HumanVMachine} (a, c, and e), show broad agreement with results obtained from human annotations. Furthermore, it should be noted the human annotations started when annotators judged that defects could be reliably marked. However, the model was capable of producing defect counts at significantly earlier times consistent with the observed scaling.
\onecolumngrid

\twocolumngrid
The spatial distribution of the defects can also be studied. The XY model makes definitive predictions for the spin-spin correlation length\cite{yurke_coarsening_1993}. If the YOLO detections can accurately resolve the spatial distribution of the defects, then measuring the average defect nearest-neighbor distance would allow for a high-resolution test of the XY predictions. The accuracy of the nearest neighbor distance is demonstrated in  Figure~\ref{fig:HumanVMachine} (b, d, and f). Though there appears to be a systematic bias, where the YOLO detections are, on average, farther apart than the hand-labelled defects, the important dynamics are captured by the \textit{scaling} of the nearest neighbor distance with time, which is resolved by the slope. As the the slope of both methods are consistent, this gives confidence for using the YOLO method for spatial analysis.

\begin{figure}
  \includegraphics[width=\linewidth]{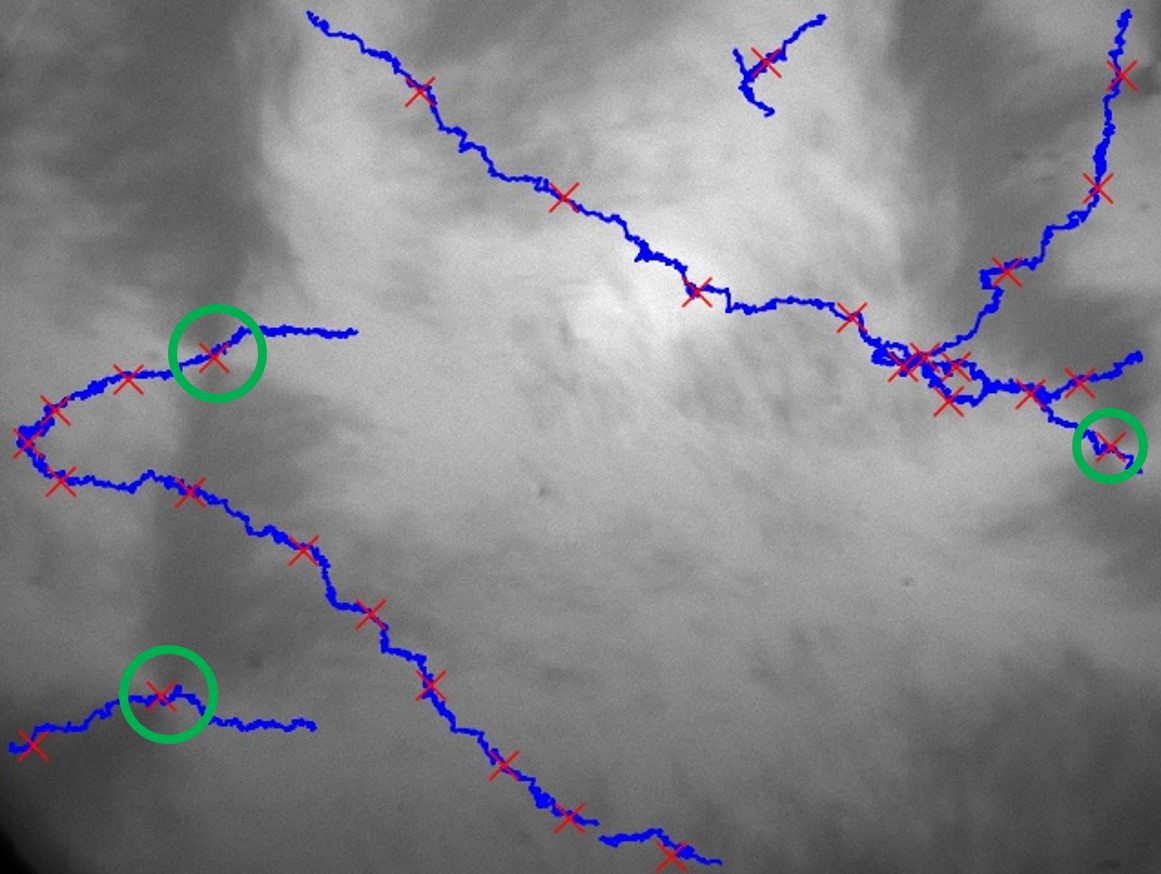}
  \caption{Computer tracked paths are represented by lines. Human annotations are represented by X's. The defects being tracked are circled. Not all tracked defects were present in a single frame, resulting in there being more tracks than circled defects. Some false detections were made, however tracks with few defect detections were omitted for clarity. }
  \label{fig:tracks} 
\end{figure}

Another application of the YOLO model to these systems is in measuring the dynamics of isolated defects.
Reliable defect tracking requires the model to be consistently capable of precisely locating defects over a larger number of frames -- a common yet challenging goal in the machine learning paradigm. We make use of the Trackpy Python module, a package of functions specializing in particle tracking, to link identified objects through consecutive images over time. The end result is a linked path for each defect, as seen in Figure~\ref{fig:tracks}.

To numerically rate the performance of our defect tracking, we compare the track to human annotated defect locations. Error is calculated by taking the root mean square of the distance in pixels between each human annotated defect location and the nearest neighbour path. For the test case in Fig. \ref{fig:tracks}, using the best performing model, we found the error to be 1.03 pixels. This shows that our machine learning pipeline is capable of tracking objects to a similar quality as a human, making it a viable method for high precision automation.

\subsection{Computational Performance}

Running the model on 1104x800 images takes approximately 0.07 seconds per image with an 8 second startup overhead on a 2017 GeForce$^{\tiny{\text{\textregistered}}}$ GTX 1080 GPU. Using an i7-7700K CPU, the model took 4.62 seconds per image with a 10 second startup overhead. When trained on the aforementioned GPU, it took 0.51 seconds per iteration using a batch size of 8 with a 12 second startup overhead, or approximately 0.064 seconds per image. When trained on the CPU, the time per image was approximately 3 seconds.   
This demonstrates the viability of using a YOLO model for rapid image data analysis, especially when used in conjunction with a modern GPU. Using a GPU, a model trained for 40 epochs on a training-set of 1000 images takes approximately 1-1.2 hours.
Training an identical model on the CPU is estimated to take between 20-40 hours, however this was not explicitly measured.

\section{Results and Discussion}
We examined the viability of using simulated images to train a modern machine learning, object detection algorithm for use in small scale applications. We demonstrate a general methodology for creating diverse training data that results in viable models with predictive power. By pairing a randomization process with the injection of characteristic experimental noise, we were able to build viable training data from simple computational simulations. 

It was found that a model trained on unmodified simulated images produced a model that performed poorly on experimental images. After increasing the diversity of simulated images via our general modification pipeline, it was found that model performance was greatly improved on experimental images, with mAP score peaking at 0.818 from a raw score of .02, with a corresponding peak F1 score of 0.811.
The model resulted in comparable spatial and number resolution to the human annotations, with significant decrease in the time-per-frame (faster analysis), resulting in a dramatic increase in the time-resolution (more frames analyzed). Additionally, the model was able to out-perform human analysis in high defect density images, which significantly supplemented the usable data.

When used in conjunction with Trackpy, the model was able to track defects with an error of 1.03 pixels compared to human annotations. This could potentially be generalized to other non-trivial targets, such as active-matter nematic defects\cite{GiomiDefectdynamicsactive2014,DeCampOrientationalordermotile2015} or even tracking biological systems such as cells\cite{MeijeringTrackingcelldevelopmental2009}. This method is fast, accurate, and easily trainable on new object types, making it a useful and versatile method for video data analysis.
\section{Acknowledgments}
This work was supported by the Soft Materials Research Center under NSF MRSEC Grant~DMR-1420736 and  by NASA Grant~NNX-13AQ81G.


%
\bibliographystyle{apsrev4-1}
\bibliography{references-edit}
\end{document}